\documentclass[aip,jcp,preprint]{revtex4-1}

\usepackage{graphicx}
\usepackage{amsmath, amsthm, amssymb}
\usepackage{textcomp}
\usepackage{color}
\usepackage{dcolumn}
\usepackage{multirow}
\usepackage{bm}

\begin{document}

\title{Laser-induced molecular alignment in the presence of chaotic rotational dynamics}

\author{Johannes Flo\ss}
\author{Paul Brumer}
\affiliation{Chemical Physics Theory Group, Department of Chemistry, and Center for Quantum Information and Quantum Control, University of Toronto, Toronto, ON M5S 3H6, Canada}
\date{\today}

\begin{abstract}
Coherent control of chaotic molecular systems, using laser-assisted alignment of sulphur dioxide (SO$_2$) molecules in the presence of a static electric field as an example, is considered.
Conditions for which the classical version of this system is chaotic are established, and the quantum and classical analogs are shown to be in very good correspondence.
It is found that the chaos present in the classical system does not impede the alignment, neither in the classical nor the quantum system.
Using the results of numerical calculations, we suggest that laser-assisted alignment is stable against rotational chaos for all asymmetric top molecules.
\end{abstract}


\maketitle


\section{Introduction}

Coherent control, applied successfully to steer many quantum mechanical processes, uses the quantum nature of the light-matter interaction to actively control atomic and molecular processes~\cite{shapiro12,shapiro03b,rice00,rice01,bergmann98,rabitz00}.
The essence of coherent control is that quantum coherence effects, which are absent in classical molecular dynamics, can be manipulated by externally controllable parameters like the frequencies and phases of a laser pulse incident on the target system.
A major aim of coherent control is the steering of complex chemical reactions towards a desired product.
Under these circumstances one can anticipate encountering chaotic motion, i.e. extreme sensitivity of the classical phase space trajectories to the initial conditions~\cite{brumer81,brumer88,zhao05}.
A better understanding of the coherent control of chaotic molecular systems is thus of importance to extend coherent control of molecular processes.

Control of quantum chaos has received some attention in the past~\cite{gong05}, using simple model systems for which the underlying classical dynamics are chaotic. 
These include the control of quantum chaotic diffusion~\cite{gong01,gong01a} -- demonstrated in a recent experiment~\cite{bitter16a} -- , branching ratios of photodissociation~\cite{abrashkevich02}, or the translational dynamics of few-level atoms in optical lattices~\cite{gong02}.

Laser-induced molecular alignment is a well established coherent control technique~\cite{stapelfeldt03,seideman05,ohshima10,fleischer12}, where a short, non-resonant laser pulse acts as a kick to a molecule, exciting a rotational wave packet via multiple coherent Raman-type interactions~\cite{zon75,friedrich95,friedrich95b}.
By combining several laser pulses and varying delays, polarization or intensity, the control can be extended to 3D-alignment~\cite{lee06,viftrup07}, unidirectional rotation~\cite{karczmarek99,fleischer09,zhdanovich11}, or selective excitations in mixtures~\cite{fleischer06,floss12}.
Laser-assisted control of molecular rotations have found many applications ranging, for example, from attosecond spectroscopy~\cite{velotta01,itatani05,wagner07} over the control of molecular collisions with atoms~\cite{tilford04} and surfaces~\cite{kuipers88,tenner91} to the steering of chemical reactions~\cite{larsen99}.

One intriguing aspect of laser-assisted control of molecular rotations is that, in many cases, it can be understood using a simple classical picture~\cite{karczmarek99,averbukh01,khodorkovsky11}.
We might, therefore, expect that a coherent control scheme that is relatively well understood classically is likely to be affected by classical chaos, if the latter is manifest in the system.

In this paper, we provide a theoretical study of laser-induced alignment in the presence of chaotic rotational dynamics.
The objects of study are sulphur dioxide (SO$_2$) molecules in a static electric field, interacting with a femtosecond laser pulse that induces molecular alignment.
SO$_2$ is an asymmetric top rotor which, in free space, has complicated but integrable dynamics.
The interaction of the molecular dipole with the static field converts the rotational dynamics from complicated to chaotic.
Below, we investigate how the chaotic dynamics affect the possibility of alignment, and find that laser-induced alignment of SO$_2$ is robust against rotational chaos.
Our simulations show that this robustness is likely a generic effect, and thus that laser-induced alignment of asymmetric top molecules is stable against rotational chaos.
Note that laser-assisted alignment of asymmetric top molecules in the presence of a static electric field has been studied before~\cite{filsinger09,hansen11};
however, the conditions in those studies did not give rise to chaotic dynamics.

This paper is structured as follows.
In Sec.~\ref{sec.methods}, we introduce the model and describe the details of our numerical methods.
Section~\ref{sec.chaos} is devoted to an investigation of chaos in asymmetric top molecules interacting with an electric field.
Although there are a few earlier works that indicate the presence of chaos in this system~\cite{barrientos95,grozdanov96,rahim05,arango08}, they concentrate on slightly different systems or aspects of the dynamics.
Hence, it was necessary to establish a better background of our object of study.
The results for laser-induced alignment of SO$_2$ in a static electric field are presented in Sec.~\ref{sec.alignment} and are generalized for other molecules in Sec.~\ref{sec.atall}. 
In this section, we also present simulations that indicate a possible origin of the chaotic motion.
Finally, in Sec.~\ref{sec.conclusion}, we discuss the results and conclude.


\section{\label{sec.methods}Model and numerical methods}

\subsection{Model}

\begin{figure}
\includegraphics[width=3.375 in]{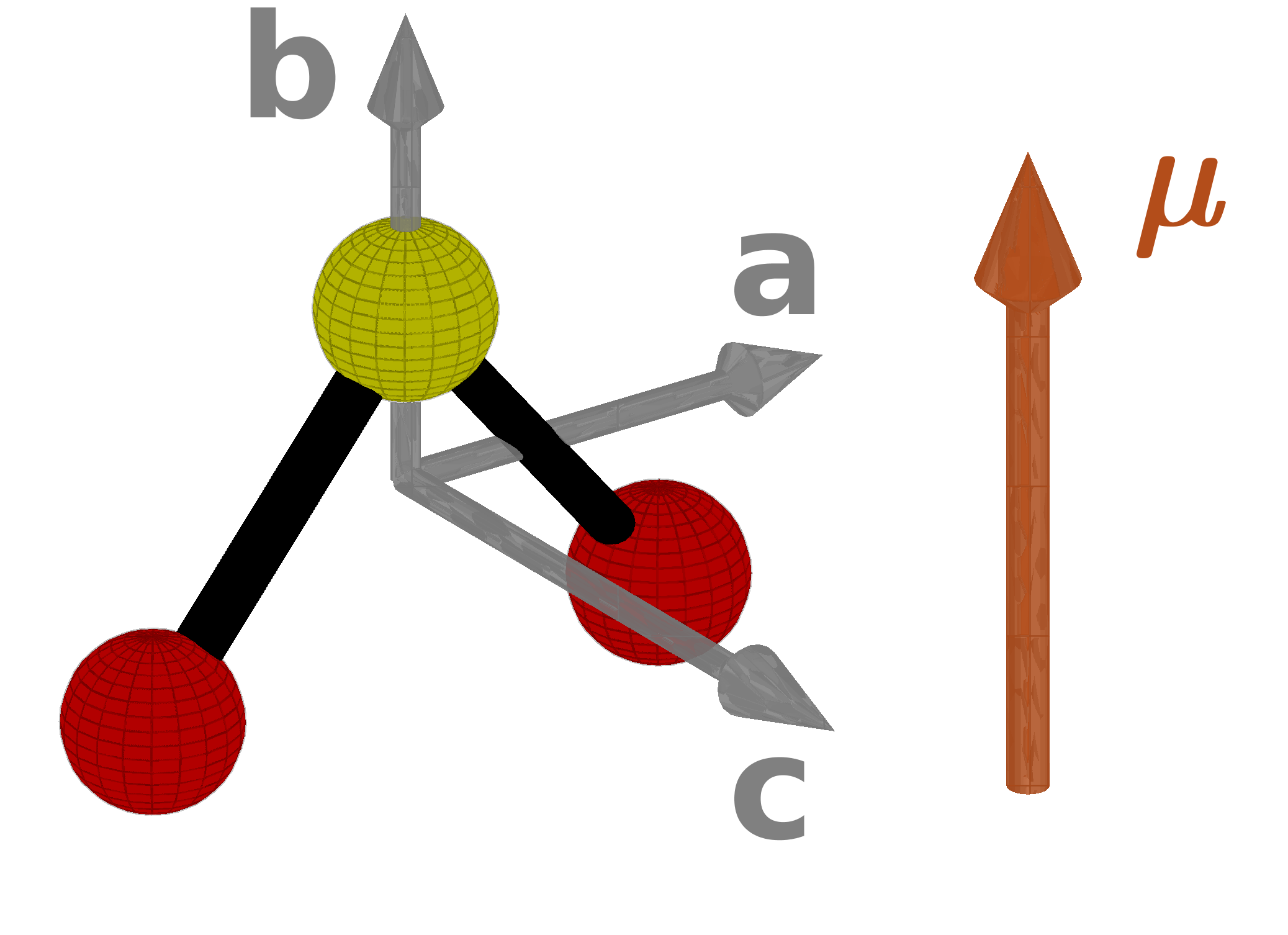}
\caption{
\label{fig.molecule}
Sketch of a sulphur dioxide (SO$_2$) molecule with the directions of the principle moment of inertia axes $a$, $b$, and $c$, and the dipole moment $\bm{\mu}$.
}
\end{figure}

As an example for an asymmetric top molecule, we consider sulphur dioxide in its most abundant isotopologue $^{32}$S$^{16}$O$_2$.
A sketch of the molecule with its principle moment of inertia axes is shown in Fig.~\ref{fig.molecule}.
As customary, the axes are labeled according to increasing moment of inertia, $I_a<I_b<I_c$.
In particular, we use the values $I_a=1.38031\cdot10^{-46}~\text{kg m}^2$, $I_b=8.13077\cdot10^{-46}~\text{kg m}^2$, and $I_c=9.53361\cdot10^{-46}~\text{kg m}^2$.
The molecular electric dipole, $\mu=1.66~\text{D}$, points along the $b$-axis.
For our investigations we consider SO$_2$ as a rigid molecule.
Note that due to the zero nuclear spin of the oxygen atoms, the rotational wave functions have to be symmetric with respect to a rotation around the $b$-axis.

The asymmetry is often characterized by an asymmetry parameter $\kappa=(2B-A-C)/(A-C)$, where $A=\hbar^2/(2I_a)$, $B=\hbar^2/(2I_b)$, $C=\hbar^2/(2I_c)$ are the rotational constants.
$\kappa$ can lie between $-1$ (prolate symmetric top) and $+1$ (oblate symmetric top), and is 0 at the highest asymmetry.
For SO$_2$, it is $\kappa=-0.94$, and thus this molecule is often considered as a near-prolate asymmetric top.

Although SO$_2$ is not a generic asymmetric top rotor due to its symmetry plane, it offers several advantages.
First, laser-induced alignment of SO$_2$ has been the object of recent experimental studies~\cite{lee06a,tenney16}, allowing us to compare our results with experiment.
Second, being a near-prolate rotor, it provides us with the option to study the effects of chaos on rotational revivals as well.
And finally, the zero nuclear spin of the $^{16}$O-atoms reduces the number of allowed symmetries for the rotational wave functions, simplifying the study.

\begin{figure}
\includegraphics[width=3.375 in]{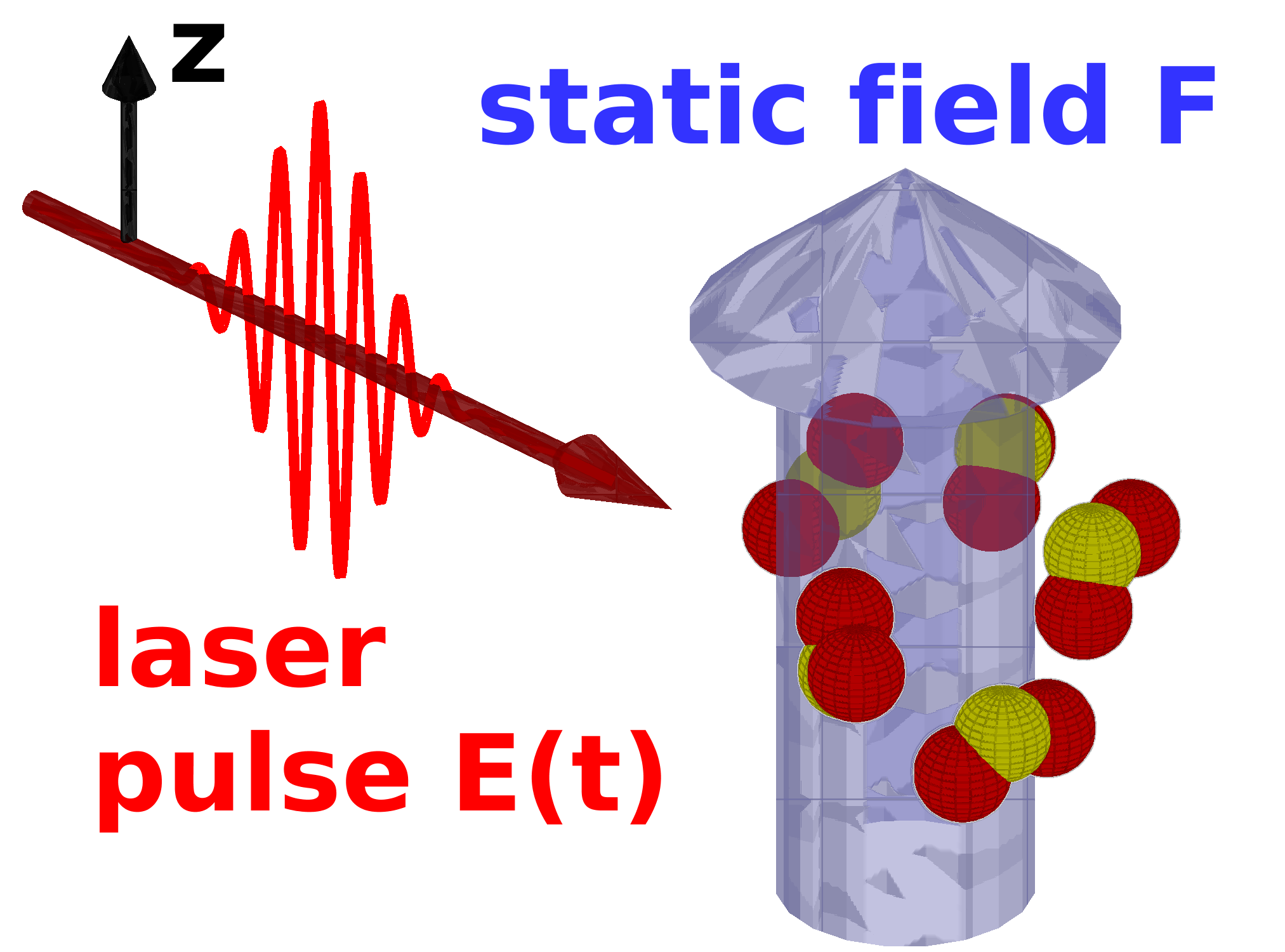}
\caption{
\label{fig.setup}
A sketch of the considered scenario.
SO$_2$ molecules rotate in a static electric field $\mathbf{F}$.
The direction of the static field defines the laboratory $z$-axis.
An off-resonant femtosecond laser pulse imparts a rotational kick to the molecules.
The laser-pulse is linearly polarized, parallel to the static field.
}
\end{figure}

The considered setup is shown in Fig.~\ref{fig.setup}.
The molecules rotate in an electric field $\mathbf{F}$, which defines the laboratory $z$-axis.
A femtosecond laser pulse -- with its polarization direction parallel to the static field -- is applied.
The laser is far detuned from any molecular excitations, and its intensity is below the ionization threshold.
A typical laser pulse is 50-100~fs long with a peak intensity of 10-50~TW/cm$^{2}$.
The laser field affects the molecular rotation via Raman-type interactions~\cite{zon75,friedrich95,friedrich95b}.
The electric field of the pulse induces anisotropic molecular polarization, interacts with the induced dipole, and tends to align the most polarizable molecular axis along the laser polarization direction.
For SO$_2$, the most polarizable axis is the $a$-axis.
An ultrashort laser pulse acts like a kick exciting molecular rotation, and the alignment is observed under field-free conditions after the pulse is over~\cite{ortigoso99,seideman99a,underwood05}.
A typical measure for the degree of the alignment of a molecular axis is the square of this axis' direction cosine with respect to the laser polarization axis, called the alignment factor.
It is unity for maximal alignment, zero when the two axes are perpendicular, and $1/3$ in the isotropic case.
References~\onlinecite{stapelfeldt03,seideman05,ohshima10,fleischer12} provide detailed reviews on laser-induced molecular alignment.

The Hamiltonian for the molecule and fields is given as
\begin{equation}
H=H_{\mathrm{rot}} + V_{\mathrm{st}} + V_{\mathrm{las}} (t) \,,
\label{eq.hamil}
\end{equation}
where the interaction $V_{\mathrm{st}}$ between the permanent dipole and the static field is constant in time, and the laser-induced potential $V_{\mathrm{las}}$ is time-dependent.
The rotational Hamiltonian $H_{\mathrm{rot}}$ is given as
\begin{equation}
H_{\mathrm{rot}} = \frac{J_a^2}{2I_a} + \frac{J_b^2}{2I_b} +\frac{J_c^2}{2I_c}  \,,
\label{eq.HrotGen}
\end{equation}
where $J_a$, $J_b$, and $J_c$ are the projections of the angular momentum on the molecular $a$-, $b$- and $c$-axis, respectively.
For the interaction potentials, we describe the orientation of the molecule in space by using Euler angles $(\theta,\phi,\chi)$, where we apply the $y$-convention~\cite{goldstein01}.
Also, for the body-fixed coordinate system we use the order $(c,a,b)$, such that the polar angle $\theta$ is the angle between the molecular dipole and the laboratory $z$-axis.
The field-dipole-interaction is given as
\begin{equation}
V_{\mathrm{st}} = - \mathbf{F}\cdot\bm{\mu} + F\mu = F\mu(1-\cos\theta) \,.
\label{eq.Vst}
\end{equation}
The term $F\mu$ is added to set zero energy as the minimum of the potential.
For the laser-induced potential, applying the usual average over the fast oscillations of the electric field~\cite{seideman05} provides an effective potential
\begin{equation}
V_{\mathrm{las}}(t)=-\frac{\varepsilon^2(t)}{4}\left(
\alpha_{aa} |\langle a|z\rangle|^2 + \alpha_{bb} |\langle b|z\rangle|^2 + \alpha_{cc} |\langle c|z\rangle|^2
\right) \,.
\label{eq.Vlas}
\end{equation}
Here, $\varepsilon^2(t)$ is the intensity envelope of the laser pulse, $\alpha_{ii}$ is the molecular polarizability along axis $i$, and $|\langle i | z \rangle|^2$ are the squares of the direction cosines with respect to the $z$-axis.
They are given in terms of Euler angles as
\begin{subequations}
\begin{align}
|\langle a | z \rangle |^2 =& \sin^2\theta \sin^2\chi \\
|\langle b | z \rangle |^2 =& \cos^2\theta \\
|\langle c | z \rangle |^2 =& \sin^2\theta \cos^2\chi \,.
\end{align}
\label{eq.direcos}
\end{subequations}
For the molecular polarizabilities we use the values from Ref.~\onlinecite{lukins85}:
$\alpha_{aa}=5.42~\text{\AA}^3$, $\alpha_{bb}=3.41~\text{\AA}^3$, and $\alpha_{cc}=2.84~\text{\AA}^3$.

It is useful to introduce the ratio $\Phi=2\mu F/E_{\mathrm{tot}}$, i.e. the ratio of the depth of the static field potential to the total energy.
In absence of the laser field, the dynamics depend only on the ratio $\Phi$, not on the absolute strength of the static field (up to a global scaling of the time).

\subsection{Quantum mechanical treatment}

It is convenient to use symmetric top wave functions $|JKM\rangle$ as basis for the quantum mechanical studies of the system.
Here, the quantum number $J$ is the total angular momentum, $K$ is its projection on a molecule-fixed axis, and $M$ its projection on a space-fixed axis.
It is beneficial to choose the $b$-axis as the molecule-fixed axis:
Thus, the constraint of even rotations around $b$, imposed by the nuclear spin of the oxygen atoms, is fulfilled by allowing only even $K$.
The laboratory $z$-axis is chosen as the space-fixed axis, so that $M$ is conserved under both interactions $V_{\mathrm{st}}$ and $V_{\mathrm{las}}$.

The non-zero matrix elements of the rotational Hamiltonian for the symmetric top wave functions are given by
\begin{subequations}
\label{eq.Hrot}
\begin{align}
\langle J K M|& H_{\mathrm{rot}} | J K M \rangle =
\frac{C+A}{2}\left[J(J+1)-K^2\right] + B K^2 \\
\langle J K M|& H_{\mathrm{rot}} | J K\pm2 M \rangle = 
\frac{C-A}{4} \cdot f(J,K\pm1)
\end{align}
\end{subequations}
with
\begin{equation}
f(J,K)=\sqrt{(J-K)(J-K+1)(J+K)(J+K+1)} \,,
\end{equation}
where we recall that $A=\hbar^2/(2I_a)$, $B=\hbar^2/(2I_b)$, and $C=\hbar^2/(2I_c)$ are the rotational constants.
It is beneficial to express the interaction potentials in terms of the Wigner rotation matrices $D_{pq}^{(s)*}$ as
\begin{equation}
V_{\mathrm{st}} = F\mu \left(1-D_{00}^{(1)*}\right)
\end{equation}
and
\begin{equation}
V_{\mathrm{las}} = \frac{\varepsilon^2(t)}{4} \Big[
\frac{2\alpha_{bb}-\alpha_{aa}-\alpha_{cc}}{3}D_{00}^{(2)*} 
 - \frac{\alpha_{aa}-\alpha_{cc}}{\sqrt{6}} \left( D_{02}^{(2)*} + D_{0-2}^{(2)*} \right)
\Big]
\label{eq.coupLaser}
\end{equation}
and use the relation~\cite{brown03}
\begin{equation}
\langle JMK | D_{pq}^{(s)*} |J'M'K'\rangle = (-1)^{M-K} \sqrt{(2J+1)(2J'+1)}
\left(\begin{array}{ccc}J&s&J'\\-M&p&M'\end{array}\right)
\left(\begin{array}{ccc}J&s&J'\\-K&q&K'\end{array}\right) \,,
\label{eq.Wigner}
\end{equation}
where the large brackets are the Wigner~3-j symbols.
The static potential thus couples states with $\Delta J=\pm1$ and $\Delta K=0$, whereas the laser interaction couples $\Delta J=0,\pm1,\pm2$ and $\Delta K=0,\pm2$.

Since the laser field is turned on only for a short time, we will use the eigenfunctions $|\varphi_n\rangle$ of $H_0\equiv H_{\mathrm{rot}}+V_{\mathrm{st}}$ as basis functions.
They are obtained as linear combinations $|\varphi_n^{(M)}\rangle=\sum_{JK} c_{JK}^{(M)} |JKM\rangle$ of the symmetric top eigenfunctions by numerically diagonalizing $H_0$.
Here, we use the fact that both $H_{\mathrm{rot}}$ and $V_{\mathrm{st}}$ (and also $V_{\mathrm{las}}$) conserve $M$, and treat $M$ as a parameter.
The remaining quantum numbers are included in $n$.
The corresponding eigenvalues of $H_0$ are denoted as $E_n^{(M)}$.
For the ease of reading, we drop the superscript $(M)$ below.

Note that for a generic asymmetric top without an external field, the rotational eigenfunctions divide into four groups according to their symmetry with respect to rotations around the principal axes.
However, the electric field destroys the symmetry for rotations around the $a$- and $c$-axes, and the zero nuclear spin demands that only eigenfunctions symmetric with respect to a rotation around the $b$-axis are allowed (this is ensured by using only even $K$);
thus all $|\varphi_n\rangle$ are of the same symmetry.

To calculate the effect of the laser pulse, we numerically solve the time-dependent Schr\"odinger equation,
\begin{equation}
i\hbar\frac{\partial |\Psi(t)\rangle}{\partial d t} = \left[H_0+V_l(t)\right] |\Psi(t)\rangle \,.
\end{equation}
Expressing $|\Psi(t)\rangle=\sum_n C_n(t)\exp(-iE_n t/\hbar) |\varphi_n\rangle$ as a linear combination of the eigenfunctions of $H_0$ gives a set of coupled differential equations for the expansion coefficients $C_n(t)$:
\begin{equation}
i\hbar\frac{\partial C_n(t)}{\partial t} = \sum_m C_m(t) e^{-i(E_m-E_n)t/\hbar} \langle \varphi_n | V_{\mathrm{las}} | \varphi_m\rangle \,.
\label{eq.tdse}
\end{equation}
Equations~\eqref{eq.tdse} are solved numerically, using approximately 3000 states as basis set for each $M$ (corresponding to $J_{\text{max}}\approx75$), with $|M|\le8$.
We checked for convergence of the results with respect to the basis size.
The coupling elements $\langle \varphi_n | V_{\mathrm{las}} | \varphi_m\rangle$ are obtained by expressing $|\varphi_n\rangle$ in the symmetric top basis and using Eqs.~\eqref{eq.coupLaser} and~\eqref{eq.Wigner}.

The expectation value of the alignment factor $\langle A_a \rangle$ for the $a$-axis is given as
\begin{align}
\langle A_a \rangle =& \langle \sin^2\theta \sin^2\chi \rangle \nonumber\\
=& \langle \Psi(t)|\left[ \frac{1}{3} -\frac{1}{3} D_{00}^{(2)*} - \frac{1}{6} \left( D_{02}^{(2)*} + D_{0-2}^{(2)*} \right)\right]|\Psi(t)\rangle \,.
\end{align}
To include thermal effects, we do ensemble averaging; i.e., we calculate the alignment factor for each initial state and add the results, weighted by the respective Boltzmann factor of the initial state.

\subsection{\label{sec.methodsClass}Classical treatment}

Using Euler angles, as in Eqs.~\eqref{eq.Vst} and~\eqref{eq.direcos}, is very inconvenient for numerical purposes, since it causes singularities in the equations of motion.
Instead, for the classical numerical treatment we utilize so-called Euler parameters $\mathbf{q}=(q_0,q_1,q_2,q_3)$, related to the Euler angles via~\cite{goldstein01}
\begin{subequations}
 \begin{align}
 q_0=&\cos\left(\frac{\phi+\chi}{2}\right)\cos\left(\frac{\theta}{2}\right) \\
 q_1=&\sin\left(\frac{\phi-\chi}{2}\right)\sin\left(\frac{\theta}{2}\right) \\
 q_2=&\cos\left(\frac{\phi-\chi}{2}\right)\sin\left(\frac{\theta}{2}\right) \\
 q_3=&\sin\left(\frac{\phi+\chi}{2}\right)\cos\left(\frac{\theta}{2}\right) \,.
 \end{align}
 \label{eq.quarternion}
\end{subequations}
One can also define conjugate momenta $\mathbf{p}=(p_0,p_1,p_2,p_3)$ for the Euler angles as~\cite{arribas06}
\begin{subequations}
 \begin{align}
  p_0=&-2q_1J_a-2q_2J_b-2q_3J_c \\
  p_1=&2q_0J_a+2q_2J_c-2q_3J_b \\
  p_2=&2q_0J_b+2q_3J_a-2q_1J_c \\
  p_3=&2q_0J_c+2q_1J_b-2q_2J_a \,.
 \end{align}
 \label{eq.momenternion}
\end{subequations}
The conjugate momenta are not required for solving the equations of motion, but they will be used to calculate distances of trajectories in phase space (later below).
The static field interaction potential is given in terms of Euler parameters as
\begin{equation}
V_{\mathrm{st}}= F\mu \left(1 - q_0^2 + q_1^2 + q_2^2 - q_3^2\right) \label{eq.VsEuler}
\end{equation}
and the laser interaction potential as
\begin{equation}
V_{\mathrm{las}}= - \frac{\varepsilon^2(t)}{4}
\Big[ \alpha_{aa} \left(2q_2q_3-2q_0q_1\right)^2 
+ \alpha_{bb} \left(q_0^2-q_1^2-q_2^2+q_3^2\right) +  \alpha_{cc} \left(2q_1q_3+2q_0q_2\right)^2 \Big] \label{eq.VlEuler} \,.
\end{equation}
Following the procedure described in~\cite{shivarama04}, the equations of motion for the rigid rotor problem are then
\begin{subequations}
\label{eq.eom1}
\begin{align}
\frac{\partial \mathbf{h}}{\partial t} =& - \boldsymbol{\omega} \times \mathbf{h} - \frac{1}{2} \mathbf{G} \frac{\partial V}{\partial \mathbf{q}} \label{eq.eom1a}\\
\frac{\partial \mathbf{q}}{\partial t} =& \frac{1}{2} \mathbf{G}^{T} \mathbf{I}^{-1} \mathbf{h} \label{eq.eom1b} \,.
\end{align}
\end{subequations}
Here, $\mathbf{h}=\mathbf{I} \boldsymbol{\omega}$ is the angular momentum vector in the body-fixed frame, where $\mathbf{I}$ is the inertia tensor and $\boldsymbol{\omega}$ is the vector of the angular velocities.
The matrix $\mathbf{G}$ is defined as
\begin{equation}
\mathbf{G}= \left(\begin{array}{rrrr}
-q_1 & q_0 & q_3 & -q_2 \\
-q_2 & -q_3 & q_0 & q_1 \\
-q_3 & q_2 & -q_1 & q_0
\end{array}\right) \,.
\label{eq.G}
\end{equation}
Finally, $\mathbf{G}^{T}$ is the transpose of $\mathbf{G}$, and $\mathbf{I}^{-1}$ the inverse of $\mathbf{I}$.

We solve Eqs.~\ref{eq.eom1} numerically by the fourth order Runge-Kutta method, using the parameters of Dormand and Prince~\cite{press07}, with dense output produced by using the parameters of Shampine~\cite{shampine86}.

Monte-Carlo-sampling over the initial conditions was used in order to include finite temperatures, where the relative weight $f$ of one set of initial conditions $(\mathbf{q}^{(0)}, \mathbf{p}^{(0)})$ is given as 
\begin{equation}
f(\mathbf{q}^{(0)}, \mathbf{p}^{(0)})=\sin(\theta)
\exp\left[-E(\mathbf{q}^{(0)}, \mathbf{p}^{(0)})/(k_B T)\right] \,.
\end{equation}
Here, $E(\mathbf{q}^{(0)}, \mathbf{p}^{(0)})$ is the initial energy, $k_B$ is Boltzmann's constant, and $T$ is the temperature.

To calculate the phase space distance $\Delta$ between two trajectories, $(\mathbf{q},\mathbf{p})$ and $(\mathbf{q}',\mathbf{p}')$, we scale the Euler parameters and their conjugate momenta as $\mathbf{\tilde q}= \mathbf{q}\sqrt{E_{\mathrm{tot}} I_{\mathrm{tot}}}/\hbar$ and $\mathbf{\tilde p}= \mathbf{p}/\sqrt{E_{\mathrm{tot}} I_{\mathrm{tot}}}$, where $E_{\mathrm{tot}}$ is the total energy and $I_{\mathrm{tot}}=I_a+I_b+I_c$ is the total moment of inertia, and take the distance between the scaled coordinates, $\Delta=\sqrt{(\mathbf{\tilde q}-\mathbf{\tilde q}')^2+(\mathbf{\tilde p}-\mathbf{\tilde p}')^2}$.
This scaling ensures that the Euler parameters and their conjugate momenta have the same order of magnitude.

To obtain Lyapunov exponents, a measure for the time-scale of chaotic divergence, we propagate 100,000 pairs of initially close trajectories $(\mathbf{q},\mathbf{p})$ and $(\mathbf{q}',\mathbf{p'})$, and observe the difference $\Delta^{(3)}(t)=|q_3(t)-q_3'(t)|$ of the Euler parameters $q_3$ and $q_3'$.
Note that it is in general sufficient to observe the exponential divergence of only one coordinate to obtain Lyapunov exponents~\cite{} (we verified this assumption for the present system by test calculations).
To calculate the Lyapunov exponent \textit{before} the laser pulse, the initial state for the first trajectory of each pair is obtained from Monte-Carlo sampling (see above).
The initial state of the second trajectory is obtained by using the values of the first trajectory and shifting $q_3$ by $10^{-9}$, thus $\Delta^{(3)}(t=0)=10^{-9}$.
If the average (averaged over 100~ps) $\Delta^{(3)}$ after approximately 2~ns is larger than $e^{-3}$, the trajectory is considered as chaotic, and the Lyapunov exponent for this trajectory pair is calculated as $10/(t_2-t_1)$, where $t_1$ is the first time $\Delta^{(3)}$ exceeds $e^{-11}$ and $t_2$ the first time $\Delta^{(3)}$ exceeds $e^{-1}$.
Finally, we average over all calculated Lyapunov exponents.
Note that the trajectory pairs that were not found to be chaotic are not included in this average.

To calculate the Lyapunov exponent \textit{after} the laser pulse, the same procedure is used.
However, to obtain the initial coordinates for the first trajectory of each pair, we use a trajectory from a thermal ensemble, and propagate it through the laser pulse.

In principle, also the laser-induced potential can give rise to chaotic dynamics.
However, for the laser pulse parameters considered in this article, we could not observe any exponential divergence within the duration of the laser pulse.


\section{\label{sec.chaos}Chaos in the asymmetric top}

In this section, we investigate if an asymmetric top molecule in a static electric field (and in the absence of a laser pulse) displays chaotic dynamics, and under which conditions.

\subsection{Classical rotor}

\begin{figure}
\includegraphics[width=3.375 in]{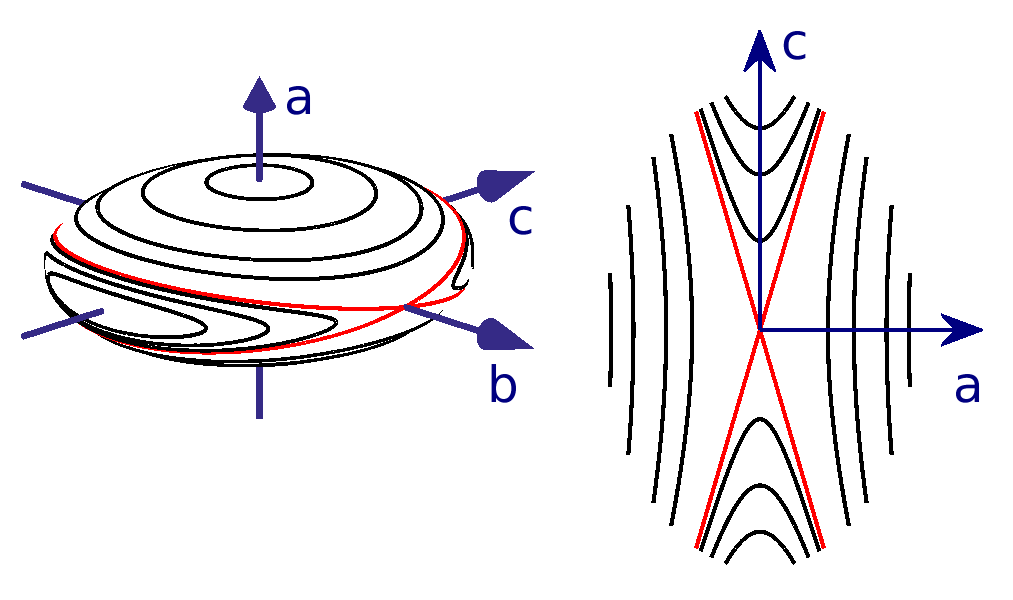}
\caption{
\label{fig.binet}
The Binet ellipsoid [see Eq.~\eqref{eq.binet}] for SO$_2$ molecules.
The black lines show intersections with the ``angular momentum sphere''~\eqref{eq.momentum} for different values of the total angular momentum $J$;
the angular momentum vector is fixed to move on such a line.
The red lines show the separatrix at $J=\sqrt{2T_{\mathrm{kin}}I_b}$.
The right picture shows the projection onto the $a,c$-plane.
}
\end{figure}

In the absence of the static field, the system becomes a textbook example of the Euler top, an integrable system.
For an Euler top, both the kinetic energy $T_{\mathrm{kin}}$ and the total angular momentum $J$ are conserved:
\begin{subequations}
\begin{align}
T_{\mathrm{kin}}=&\frac{J_a^2}{2I_a}+\frac{J_b^2}{2I_b}+\frac{J_c^2}{2I_c} \label{eq.binet} \\
J^2=&J_a^2+J_b^2+J_c^2 \label{eq.momentum} \,.
\end{align}
\label{eq.conserved}
\end{subequations}
The angular momentum vector can only move on the intersection of the ellipsoid~\eqref{eq.binet} (Binet-ellipsoid) and the momentum sphere~\eqref{eq.momentum}.
In Fig.~\ref{fig.binet}, we display the Binet ellipsoid and sample orbits of the angular momentum.
Rotations around the $a$- and $c$-axis are stable, and thus the orbits can be sorted into four groups:
(counter-) clockwise rotating around the $a$-axis as well as (counter-) clockwise rotating around the $c$-axis.
The separatrix between these groups is at momentum $J=\sqrt{2T_{\mathrm{kin}}I_b}$ (marked by red lines in Fig.~\ref{fig.binet}).
There are six stationary points for $\mathbf{J}$;
they are given by $\mathbf{J}$ being parallel to one of the principle axes.
Four of the stationary points are stable ($J=\pm J_a,\pm J_c$), and two unstable ($J=\pm J_b$).
A molecule that is initially rotating almost around the $a$- or $c$-axis will continue to do so, whereas a molecule that is initially rotating almost around the $b$-axis will undergo large oscillations of its rotation direction.
As a sidenote, for a quantum rotor also rotations around the $a$- and $c$-axes will see an inversion of the rotation, due to dynamical tunneling~\cite{harter84}.

In earlier work~\cite{barrientos95}, Barrientos~\textit{et al.} studied the dynamics of a slightly asymmetric heavy top.
The asymmetric heavy top is mathematically the same problem as an asymmetric top molecule in an electric field.
In their study they found that the slightly asymmetric heavy top shows some signs of chaos.
In the following, we conduct a similar study for SO$_2$ molecules in an electric field.
Note that, although considered as a near-prolate asymmetric top, the asymmetry of SO$_2$ is by far too large to be considered as small for the purposes of this study; thus the results might differ from Ref.~\onlinecite{barrientos95}.

We compute Poincar\'e maps as a tool to study the possible chaotic dynamics.
As the surface of section, we use the plane defined by $\theta=\pi/2$, i.e. where the molecular dipole is perpendicular to the electric field, and we plot the values of the projection of the angular momentum on the $a$- and $c$-axis for the crossings of the surface of section for decreasing $\theta$.
The angle $\phi$, the azimuthal angle of the molecular dipole, is needed to completely define the crossing;
however, since the problem is symmetric for any rotation around the laboratory $z$-axis, this angle is not significant for this study.

Note that, apart from a rescaling of the time, the classical dynamics depends only on the ratio $\Phi=2\mu F/E_{\mathrm{tot}}$ of the field-dipole interaction and the total energy, but not on their absolute values.
Also, for the chosen surface of section, only $\Phi\le2$ can be investigated; otherwise a crossing of the surface of section is impossible.

\begin{figure}
\includegraphics[width=3.375 in]{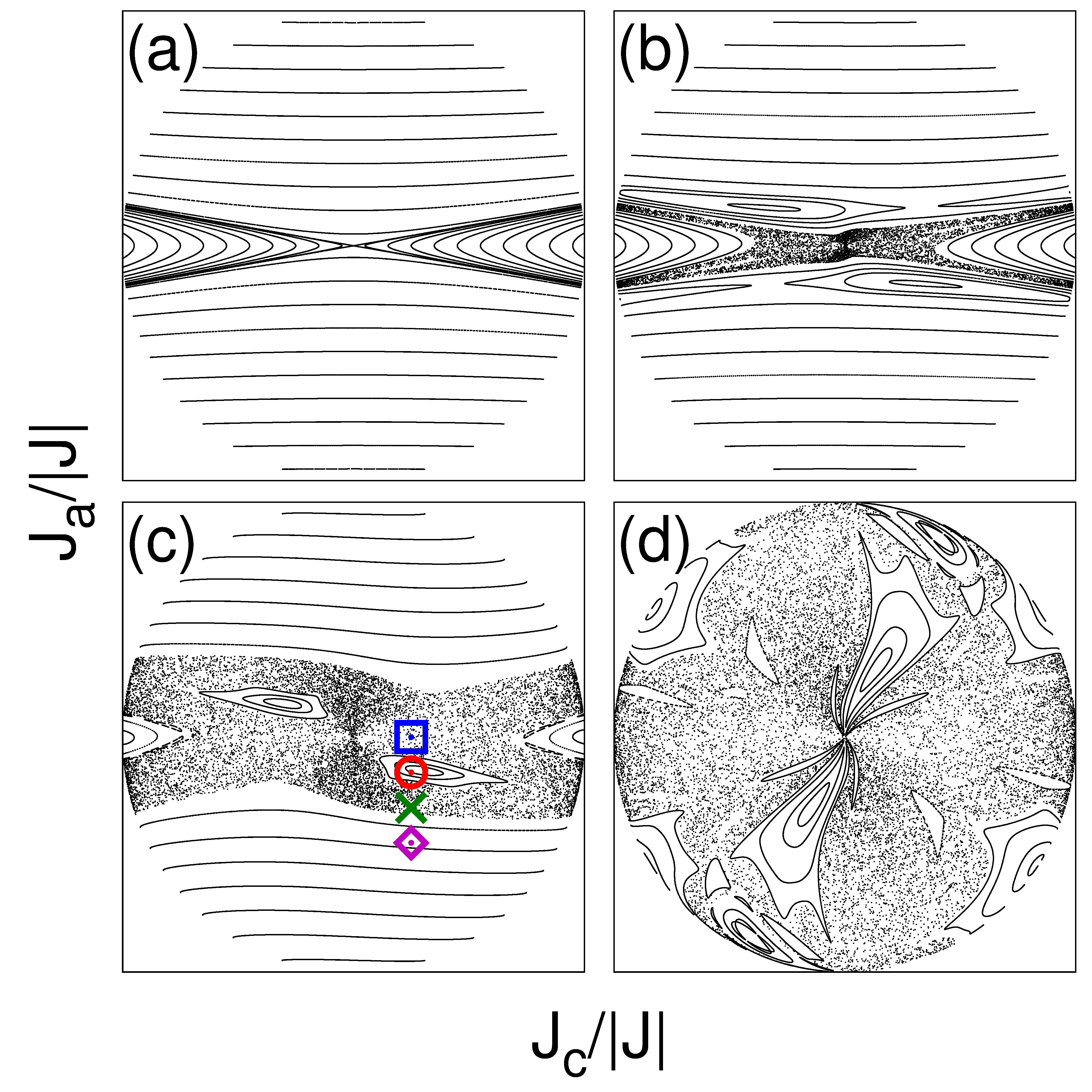}
\caption{
\label{fig.PM_pi2}
Poincar\'e maps for classical SO$_2$ molecules in an electric field.
The surface of section is defined by $\theta=\pi/2$, where $\theta$ is the angle between the molecular dipole and the electric field.
Shown are the components $J_a$ and $J_c$ of the angular momentum along the $a$- and $c$-axis at the crossing of the surface of section with decreasing $\theta$.
The coordinates are scaled to the total angular momentum $J$, thus the axes on all plots range from $-1$ to $+1$.
The ratio $\Phi=2\mu F/E_{\mathrm{tot}}$ is: (a) 0, (b) 1/50, (c) 1/5, (d) 1.
As a point of reference note that for SO$_2$ at a temperature of 10~K, $\Phi=1$ corresponds to a field strength of $\sim$150~kV/cm.
The symbols in panel~(c) mark the coordinates for the trajectories shown in Fig.~\ref{fig.trajectories}.
}
\end{figure}

The calculated Poincar\'e maps are shown in Fig.~\ref{fig.PM_pi2} for different values of $\Phi$.
The (conserved) projection $M$ of the angular momentum on the electric field axis is chosen as zero.
For the case of zero electric field, Fig.~\ref{fig.PM_pi2}~(a), only regular dynamics are present.
In fact, the location of the crossings of the surface of section are identical to the trajectories of $\mathbf{J}$ on the Binet ellipsoid, Fig.~\ref{fig.binet}.
When increasing the field strength, Fig.~\ref{fig.PM_pi2}~(b) and~(c), chaotic dynamics develop along the former separatrix.
The regions around the stable trajectories remain regular.
Finally, when the total energy is similar to the depth of the field induced potential well, Fig.~\ref{fig.PM_pi2}~(d), most parts of the surface of section display chaotic dynamics.
Note that although the larger part of the surface of section in Fig.~\ref{fig.PM_pi2}~(d) displays chaotic dynamics, this does not necessarily imply that most of the phase space is chaotic.

\begin{figure}
\includegraphics[width=3.375 in]{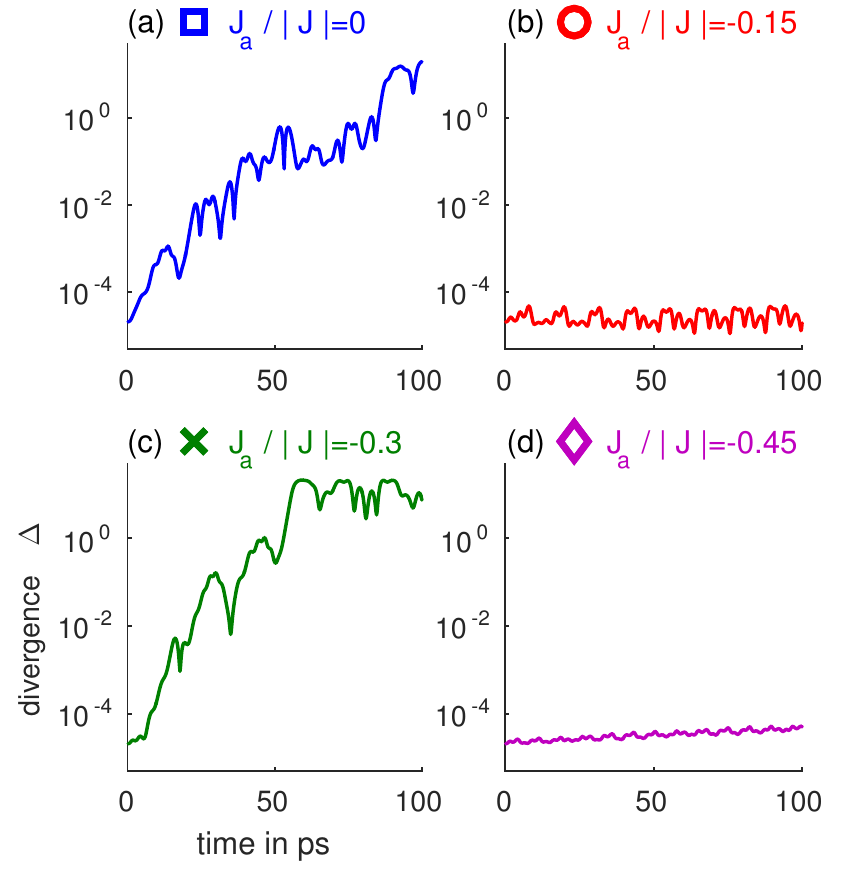}
\caption{
\label{fig.trajectories}
Time-dependence of the phase space distance $\Delta$ for sample pairs of trajectories with slightly different initial conditions.
The initial coordinates of the first trajectories are $\theta=\pi/2$, $\phi=\pi/2$, $J_c/|J|=-0.25$, and $J_a$ as displayed in the titles [colors and symbols corresponding to the symbols in Fig~\ref{fig.PM_pi2}~(c)].
For the second trajectories of each pair the initial values of all Euler parameters and angular velocities is shifted by $10^{-6}$ with respect to the first trajectory.
The other parameters are: $E_{\mathrm{tot}}=3.65~\mathrm{meV}$, $F=210~\mathrm{kV}/\mathrm{cm}$, and $M=0$, and therefore $\Phi=1/5$.
}
\end{figure}

In order to ensure that the irregular areas seen in the Poincar\'e maps do indeed display chaotic dynamics, we tested several sample trajectories for exponential divergence;
i.e. we propagated pairs of trajectories with almost identical initial conditions and examined whether the distance between them in in phase space, $\Delta$, increases exponentially over time.
These tests show that the irregular areas of the Poincar\'e maps indeed correspond to chaotic dynamics.
As an example, $\Delta$ is shown in Fig.~\ref{fig.trajectories} for four different pairs of trajectories.
Trajectories placed in a regular region [panels~(b) and~(d)] show only very slow divergence.
By contrast, the examples that represent the chaotic case [panels~(a) and~(c)] show clear exponential divergence.

One would expect that for field strengths that far exceed the total energy, the dynamics should resemble that of a harmonic oscillator, and thus be regular.
The surface of section chosen for the Poincar\'e maps shown in Fig.~\ref{fig.PM_pi2} does not allow investigation of this regime.
However, we also calculated Poincar\'e maps for other surfaces of section and verified the expected return to regular dynamics for very strong electric fields.

SO$_2$ is considered a near-prolate top, so it might be possible that being close to a symmetric top renders it less sensitive to chaotic dynamics.
We performed additional simulations with difference values for $I_b$, thus changing the asymmetry parameter $\kappa$ and found no qualitative differences.
We also changed the direction of the dipole moment vector within the molecule, but did not find any qualitative differences.

Concluding, classical chaos is present in the rotational dynamics of asymmetric top molecules in an electric field.
The chaos starts along the separatrix on the Binet ellipsoid.

\subsection{Quantum rotor}

One way to investigate quantum chaotic systems is to examine the statistical properties of their spectra~\cite{haake10}.
A typical property is the distribution $p(S)$ of the nearest neighbor spacing $S$ for the unfolded energy spectrum~\cite{bohigas84}.
It describes the fluctuations of the level spacings around the local average. 
For a regular, integrable system with conserved quantities, level-crossings are allowed.
Thus, one can expect level clustering and therefore strong fluctuations of the level-spacing around the local average.
As a result, the distribution $p(S)$ becomes Poisson-like.
On the other hand, for a chaotic system, there are less constants of motion, leading to stronger level repulsion.
As a result, the distribution $p(S)$ is expected to be Wigner-like, with $p(S=0)=0$.

\begin{figure}
\includegraphics[width=3.375 in]{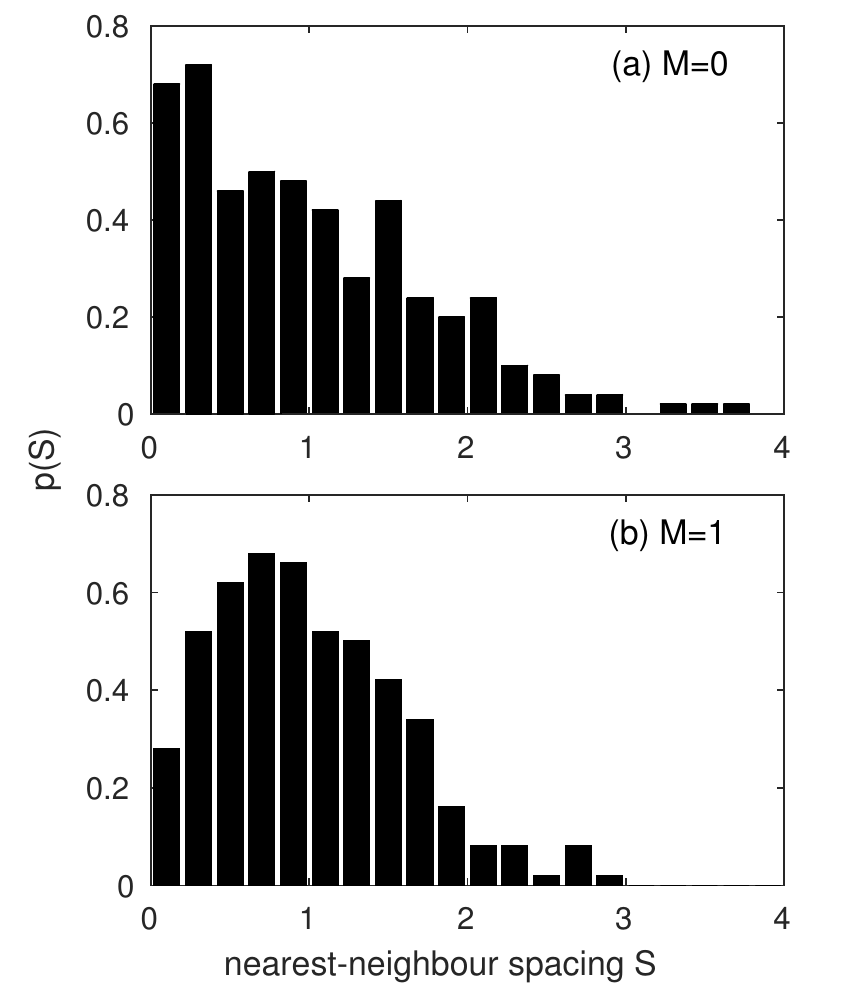}
\caption{
\label{fig.pS}
Nearest-neighbor spacing distributions for the unfolded lowest 250 rotational levels of SO$_2$ in an electric field of $F=5000~\text{kV/cm}$, for two different values of the projection quantum number $M$.
}
\end{figure}

We use the method described in Ref.~\onlinecite{bohigas84} to calculate the nearest neighbor spacing distribution $p(S)$.
Since we are interested in statistical properties of the spectra, we have to ensure that a large number of levels lie within the classically chaotic region and, to do so, we have to simulate relatively strong fields.
In Fig.~\ref{fig.pS} we show $p(S)$ for the lowest 250 rotational states of SO$_2$ in an electric field of strength $F=5000~\text{kV/cm}$, for $M=0$ and $M=1$.
The energy levels lie between $E_0=0$ and $E_{250}\approx55~\text{meV}$ (thus $2\mu F/E_{250}\approx2/3$), with a roughly constant density of states (apart from the first 50 states).
Comparing with the Poincar\'e maps, we may expect most states to be in the chaotic regime.
(Mind however that those maps do not provide a quantitative measure of the extent of chaotic and regular regions.)
For non-zero $M$, Fig.~\ref{fig.pS}~(b), we see a mixed distribution, i.e. partially Wigner-like and partially Poisson-like.
Thus, some signs of level repulsion in the classically chaotic region are present.
A different situation is found for $M=0$, Fig.~\ref{fig.pS}~(a).
Here, the distribution is almost Poisson-like, with no sign of level repulsion.
Interestingly, we did not find such a dependence on the orientation of the angular momentum for the classical system.
We repeated the calculations for SO$_2$ with a reduced moment of inertia along the $b$-axis -- thus increasing the rotor asymmetry -- but this did not lead to any qualitative changes.
In summary, the quantum asymmetric top in an external potential does show signs of chaos, although the results based on $p(S)$ are not conclusive.

There is a need for caution when using $p(S)$ to determine the degree of chaos in a system.
The level spacing distribution is a measure of the level repulsion.
Thus, all systems can display Poisson-like distributions when states of different symmetries are included simulateneously.
A chaotic system has generally less symmetries than a regular one, but it may still have some symmetries -- e.g., the system we consider here is symmetric with respect to rotations around the laboratory $z$-axis (therefore we considered states of different $M$ separately).
Then again, a regular system can also display a non-Poissonian level spacing distribution, if there are correlations between the levels;
e.g. a particle in a 2D-box, with a rational ratio of the wall lengths.
Alternative measures of quantum chaos suffer from similar ambiguities.


\section{\label{sec.alignment}Alignment}

In the preceding section, we demonstrated that a classical asymmetric top molecule in a static electric field is chaotic.
We will now investigate whether and how the chaos affects laser-assisted alignment of these molecules.
As an example, we consider rigid SO$_2$ molecules interacting with a linearly polarized laser pulse of 50~fs duration (full width at half maximum of the intensity envelope) and a peak intensity of 50~TW/cm$^2$ at an initial rotational temperature of 10~K.
The static electric field is chosen to be parallel to the laser electric field, with three different strengths: 0, 150, and 750~kV/cm.
Although some of these values might be experimentally challenging, they allow us to investigate different regimes and strengths of the chaotic dynamics.
Note also that the dynamics are scalable:
A heavier molecule at a lower temperature shows the same dynamics at lower field strengths.

For molecules that are not subject to an additional static field, it is known that a short laser pulse aligns the most-polarizable axis of the molecule, in case of SO$_2$ the $a$-axis, towards the laser polarization axis~\cite{stapelfeldt03,seideman05,ohshima10,fleischer12}.
In particular, impulsive laser alignment is known to impart a kick to the molecules, which leads to alignment of the molecules shortly after the laser pulse is over.
Subsequently, the alignment decreases again, showing quantum mechanical revivals on a longer time-scale.
Also, the time-averaged value of the alignment factor is increased.
Recall that in our setup, we choose the laser-polarization axis to be parallel to the static field, and thus the projection of the angular momentum on the electric field axis is conserved.

\begin{table}
\begin{ruledtabular}
\begin{tabular}{dddddddd}
\multicolumn{1}{c}{$F$} & \multicolumn{1}{c}{$E_{\mathrm{in}}$} & \multicolumn{1}{c}{$\Phi_{\mathrm{in}}$} & \multicolumn{1}{c}{$E_{\mathrm{fi}}$} & \multicolumn{1}{c}{$\Phi_{\mathrm{fi}}$} & \multicolumn{1}{c}{$\langle A \rangle_{\mathrm{th}}$} & \multicolumn{1}{c}{$\langle A \rangle_{\mathrm{max}}$} & \multicolumn{1}{c}{$\langle A \rangle_{\mathrm{pop}}$} \\ 
\hline
\multirow{2}{*}{0} & 1.1  & 0.0 & 4.9  & 0.0 & 0.33 & 0.64 & 0.42  \\ 
   & [1.3] & [0.0]   & [4.4] & [0.0]   & [0.33] & [0.60] & [0.40] \\
& & & & & & & \\
\multirow{2}{*}{150} & 1.5 & 0.68 & 5.3 & 0.20 & 0.33 & 0.63 & 0.41 \\ 
   & [1.7] & [0.61]  & [5.0] & [0.21]   & [0.33] & [0.61] & [0.41] \\
& & & & & & & \\
\multirow{2}{*}{750} & 2.1 & 2.5 & 5.2 & 1.0 & 0.22 & 0.53 & 0.35 \\ 
   & [2.1] & [2.4]   & [4.9] & [1.1]   & [0.22] & [0.51] & [0.34] 
\end{tabular}
\end{ruledtabular}
\caption{
\label{tab.compare}
Results of the alignment simulations for SO$_2$ molecules at different strengths $F$ of the static electric field (displayed in kV/cm).
The pulse duration is 50~fs, the peak intensity 50~TW/cm$^2$, and the initial temperature 10~K.
Presented are the initial energy $E_{\mathrm{in}}$ and final energy $E_{\mathrm{fi}}$ (in meV), as well as the ratios $\Phi_{\mathrm{in}}$ and $\Phi_{\mathrm{fi}}$ of the static potential well depth and the initial and final energy, respectively.
The last three columns display the alignment factor of the $a$-axis:
The initial (thermal) alignment $\langle A \rangle_{\mathrm{th}}$, the peak (maximal) alignment $\langle A \rangle_{\mathrm{max}}$, and the time-averaged (population) alignment $\langle A \rangle_{\mathrm{pop}}$.
The numbers in brackets are the result of a classical simulation, the other numbers are for the quantum mechanical simulation.
}
\end{table}

\begin{figure}
\includegraphics[width=3.375 in]{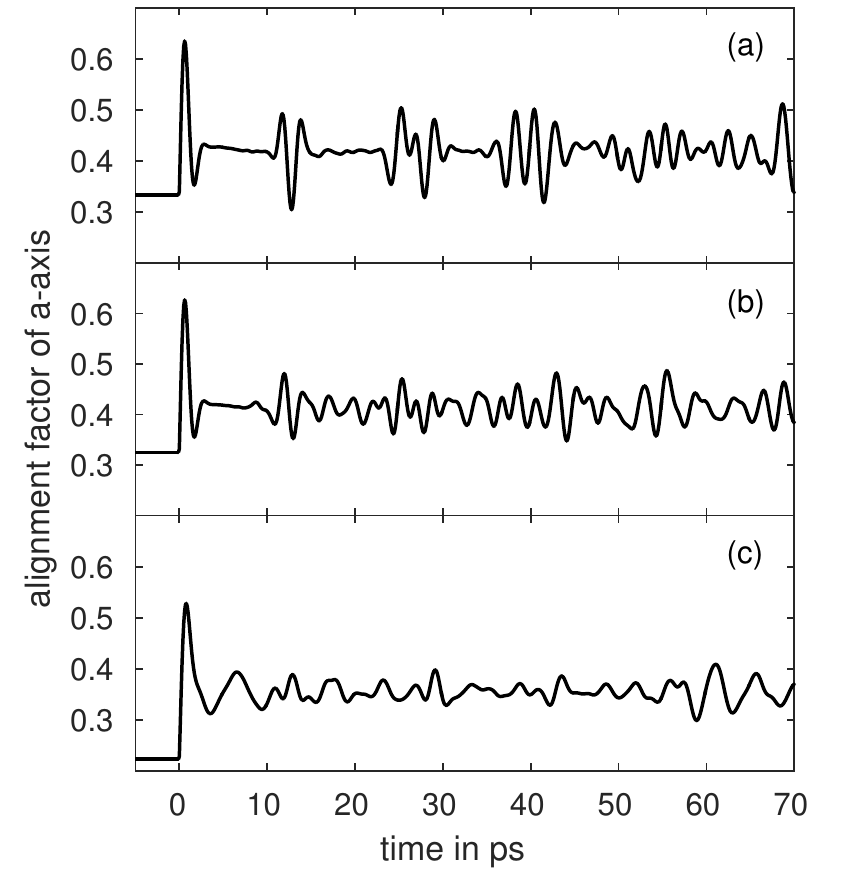}
\caption{
\label{fig.alignment}
Quantum mechanical alignment signal for the $a$-axis of SO$_2$ molecules in a static electric field at a temperature of 10~K.
At $t=0$ an aligning laser pulse is applied.
The pulse duration is 50~fs, the peak intensity 50~TW/cm$^2$, and the initial temperature of the molecules is 10~K.
The strength of the static field is (a)~0~kV/cm, (b)~150~kV/cm, and (c)~750~kV/cm.
}
\end{figure}

\begin{figure}
\includegraphics[width=3.375 in]{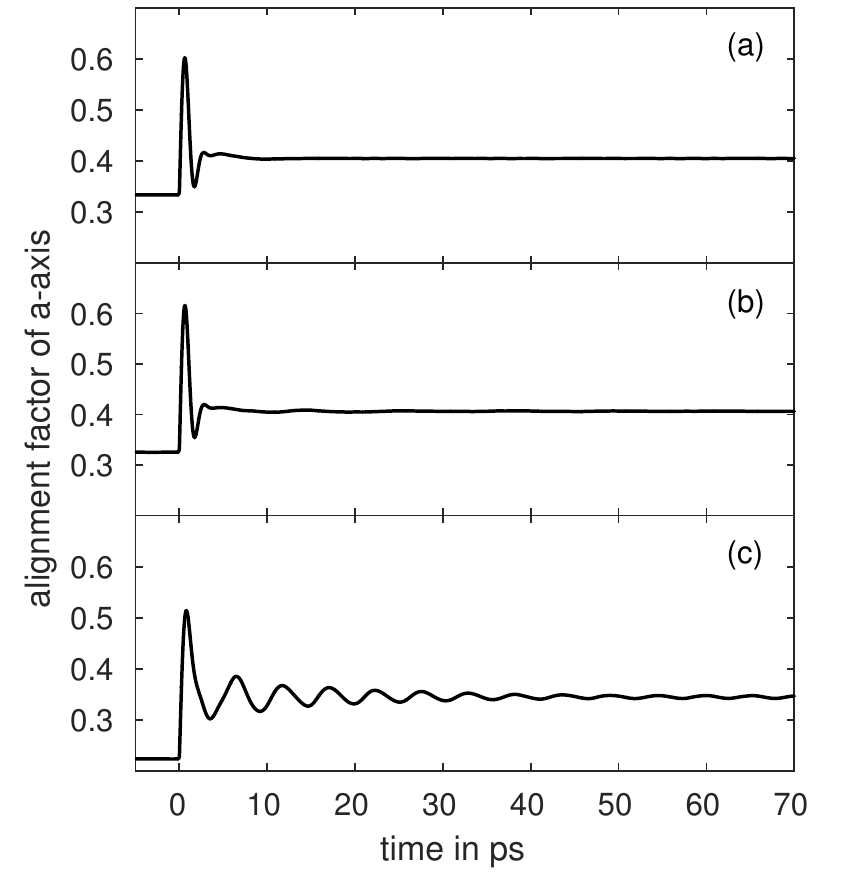}
\caption{
\label{fig.classical}
Shown are the classical alignment signals, for the same conditions as in Fig.~\ref{fig.alignment}.
}
\end{figure}

The results of the quantum and classical numerical calculations are shown in Table~\ref{tab.compare} and Figs.~\ref{fig.alignment} and~\ref{fig.classical}.
First, note the ratio $\Phi$ of the static field strength and the expectation value of the energy, which tells us about the degree of chaos in the system.
For no static field, this ratio is zero, and the dynamics are completely regular.
For the field of 150~kV/cm, the ratio is 0.68 before the pulse and 0.20 after the pulse.
Comparing with the results of Sec.~\ref{sec.chaos}, the system is chaotic before the pulse, and somewhat chaotic after the pulse.
Finally, for 750~kV/cm, the ratio changes from 2.5 to 1.0, that is from a mostly regular region (in the potential well) to a very chaotic region.
It is important to note that this ratio refers to the expectation value of the energy; the actual wave packet after the laser pulse is rather broad, covering both chaotic and regular regions for the cases of 150~kV/cm and 750~kV/cm.
It should also be noted that the energy imparted by the laser pulses is very similar for all three scenarios, between 3.1 and 3.8~meV.

Consider first the quantum results.
The alignment factor of the $a$-axis (measuring the degree of alignment of the $a$-axis, see Sec.~\ref{sec.methods}) for the three different strengths of the static field are shown in Fig.~\ref{fig.alignment}.
For all three scenarios, we can see a strong peak of the alignment factor right after the laser pulse.
Furthermore, the height of this peak is comparable for all three cases, being slightly lower for an increasing strength of the static field.
This does not appear to be related to the chaotic dynamics, but due to the competition between the two fields:
The interaction between the permanent dipole and the static field leads to an orientation of the $c$-axis and thus anti-alignment of the $a$-axis, lowering the effect of the aligning laser field.

Looking at the time-averaged alignment ($\langle A\rangle_{\mathrm{pop}}$ in Table~\ref{tab.compare}) we find the same picture.
For all three cases, the laser pulse increases the time-averaged alignment of the $a$-axis considerably.
The values are slightly lower for stronger static fields.
Again, we suggest that the reason is to be found in the anti-aligning character of the static field.

A different picture is found when looking at the revivals of the alignment signal (Fig.~\ref{fig.alignment}):
The results differ significantly for the three considered static field strengths.
For the case of no field, we find several revivals.
The most notable ones are around 13~ps, 27~ps, and 40~ps; the two latter are rather broad.
A comparison with earlier works~\cite{poulsen04,tenney16} shows that these revivals can be identified as half and full revivals of the $J$-type and the $C$-type.
The former has a period of approximately $t_J\approx26~\text{ps}$, the latter a period of $t_C\approx28~\text{ps}$.
The $J$-type revival is associated with rotations around the $b$- and $c$-axes, the $C$-type with rotations around the $c$-axis~\cite{tenney16}.
For the cases of non-zero static fields, we find that the revivals become weaker, and finally disappear, with an increasing field strength.
In additional calculations for very strong fields of 50000~kV/cm (not shown here), we found that revivals reappear, although at different times.
For such strong fields, the dynamics are well below the potential barrier and almost completely regular.
It is therefore likely that the rotational chaos is the reason for the disappearance of the revivals.

In Sec.~\ref{sec.chaos}, we showed that chaos is strongest for rotations around the $b$-axis, where even a weak static field leads to chaotic motion.
On the other hand, rotations around the $c$-axis remain regular for much stronger fields.
Thus, one naively might think the $J$-type revivals are more strongly affected by the chaos than the $C$-type revivals.
However, comparing Figs.~\ref{fig.alignment}~(a) and~(b) shows that both types are, roughly, equally affected.

In addition to the quantum mechanical calculations presented above, we also investigated the behavior of classical rotors.
The results are shown in Fig.~\ref{fig.classical} as well as Table~\ref{tab.compare} (values in brackets).
The alignment of classical rotors is seen to be only slightly less efficient than for quantum rotors.
However, there are no revivals since revivals are the result of a discrete spectrum.
Most importantly, the classical rotors show exactly the same reaction to the chaos as do the quantum rotors; that is, no evident effect.

In order to understand why chaos is ineffectual, we examine the time-scales involved.
The defining characteristic of chaotic dynamics is that the phase space distance $\Delta$ between two arbitrarily close trajectories grows exponentially over time as $\Delta(t)=\Delta(0) \exp(\lambda t)$.
The exponent $\lambda$, the Lyapunov exponent, is greater than zero for chaotic dynamics and its inverse, the Lyapunov time $t_{\lambda}=1/\lambda$, provides an estimate of the time-scale for the chaos to take effect~\footnote{For alternative chaos time scales see, e.g., J.~W.~Duff and P.~Brumer, J.~Chem.~Phys.~\textbf{71}, 2693 (1979).}.
As the time-scale for the alignment we use the time between the laser pulse and the first alignment peak.

We determined the ensemble averaged Lyapunov times $t_{\lambda}$ for the SO$_2$ rotational dynamics before and after the laser pulse for classical rotors.
See Sec.~\ref{sec.methodsClass} for a detailed description of our numerical procedure.
Note that only chaotic regions of the phase space are used to calculate the Lyapunov times.
The calculated $t_{\lambda}$ are shown in Table~\ref{tab.lyapunov}, along with the fraction $f_c$ of trajectories that are chaotic, as well as the time $t_{\mathrm{al}}$ of the first alignment peak.

\begin{table}
\begin{ruledtabular}
\begin{tabular}{dddddd}
\multicolumn{1}{c}{$F$} & \multicolumn{2}{c}{before} & \multicolumn{2}{c}{after} & \multicolumn{1}{c}{$t_{al}$}\\
& \multicolumn{1}{c}{$t_{\lambda}$ } & \multicolumn{1}{c}{$f_c$} &  \multicolumn{1}{c}{$t_{\lambda}$} & \multicolumn{1}{c}{$f_c$} & \\
\hline
 0  & \multicolumn{1}{c}{--} & 0.0 & \multicolumn{1}{c}{--} & 0.0 & 0.62 \\
 150 & 7.9 \pm 4.8 & 0.52 & 3.8 \pm 1.8 & 0.53 & 0.64 \\
 750 & 7.1 \pm 5.1 & 0.07 & 3.3 \pm 1.6 & 0.38 & 0.80 
\end{tabular}
\end{ruledtabular}
\caption{
\label{tab.lyapunov}
Average Lyapunov times $t_{\lambda}$ (given in ps) for SO$_2$ molecules in static electric fields of different strength $F$ (given in kV/cm), before and after interacting with a laser pulse.
The values are ensemble averaged for a thermal ensemble (before), and the ensemble created by the laser pulse from that thermal ensemble (after).
Only chaotic trajectories are used to calculate the average of $t_{\lambda}$.
Also shown are the fraction $f_c$ of trajectories that are chaotic, and the time $t_{\mathrm{al}}$ (in ps) of the alignment peak.
The laser pulse duration is 50~fs, its peak intensity 50~TW/cm$^2$.
All values are from classical calculations.
}
\end{table}

Comparing the Lyapunov times and the time of the peak of the alignment signal shows that the latter is significantly shorter than the former, for all strengths of the static field.
As a consequence, we can conclude that the rotational chaos is too slow to affect laser-assisted molecular alignment, at least for the considered parameters.

Note that the interaction potential between the laser field and the induced dipole is of a very similar character to the interaction potential between the static field and the permanent dipole, and could thus also induce chaotic dynamics.
However, for the laser pulses considered in this work, no exponential divergence was observed within the short duration of the laser pulses.

As for the wave packet revivals, they are a quantum phenomenon, and hence it might seem contradictory that they are the only aspect of molecular alignment that is affected by chaos, which is an inherently classical phenomenon.
However, revivals are generally not expected to occur in chaotic systems~\cite{tomsovic97}.
For a revival to occur, the energy levels have to be regularly spaced.
For example, using random matrix theory to calculate the level spacing distributions for a chaotic system, shows small fluctuations around a mean spacing~\cite{haake10}.
Since revivals are very sensitive to such fluctuations~\cite{floss14}, they are unlikely to occur in a chaotic system.

\section{\label{sec.atall}Can rotational chaos affect alignment at all?}

\begin{figure}
\includegraphics[width=3.375 in]{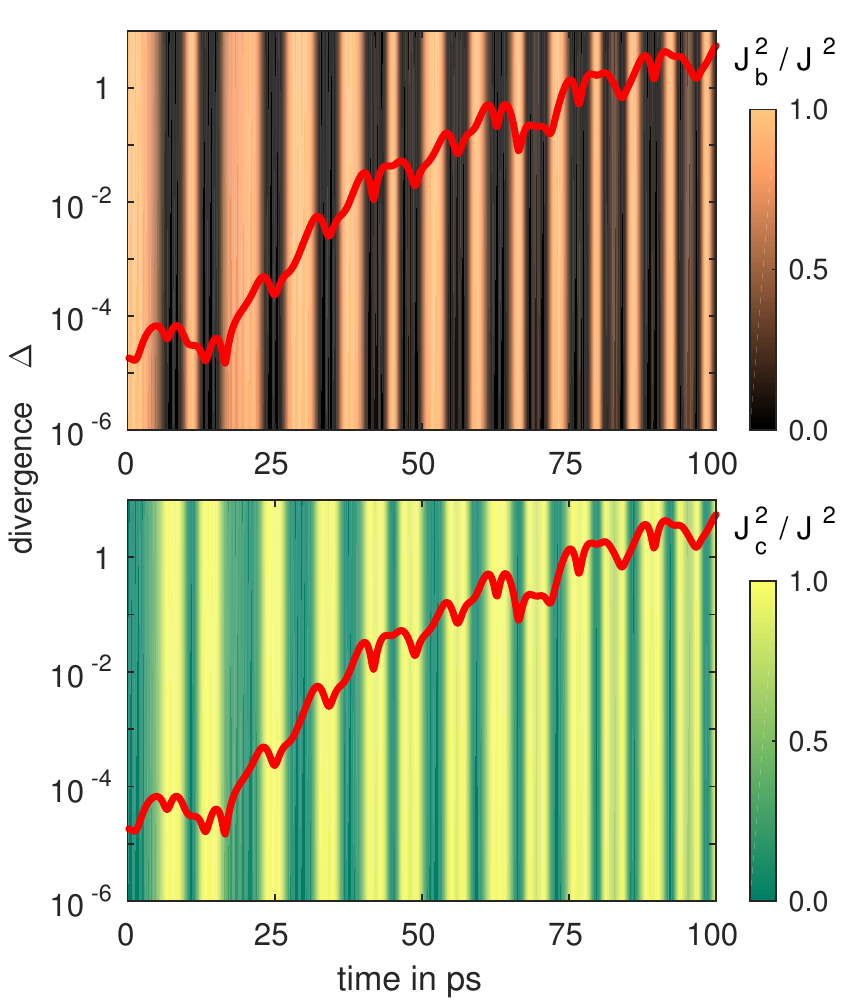}
\caption{
\label{fig.mechanism}
Time-evolution of the difference $\Delta(t)$ in phase space for an example pair of initially close rotational trajectories of SO$_2$.
The red line depicts $\Delta(t)$, and the background color displays the normalized and squared projection of the angular momentum onto the $b$-axis (upper panel) and $c$-axis (lower panel), respectively.
A static electric field of 150 kV/cm is applied, and there is no laser pulse.
}
\end{figure}

In the preceding section, we showed that laser-induced alignment of SO$_2$ molecules is robust against rotational chaos.
The key question is whether this stability is generic or only valid for the specific molecular and laser parameters used above.
As a first step towards resolving this issue we examine the mechanism of the chaos and locate the part of phase space in which the exponential divergence takes places.
Figure~\ref{fig.mechanism} shows the phase space distance $\Delta(t)$ of a pair of two initially close sample trajectories.
In the background we plot the square of the normalized projection of the angular momentum on the molecular $b$- and $c$-axis (upper and lower panel, respectively).
One can see that regions of fast growth of $\Delta$ correspond to the angular momentum preferentially pointing along the $b$-axis.
We found this behavior for the majority of investigated initial conditions, although it is not a strict rule.
This strongly suggests that the exponential divergence is connected to a rotation around the $b$-axis.

The equations of motion for this system results in a strong coupling of all momenta and coordinates.
Therefore, despite the fact that the exponential divergence is connected to a rotation around the $b$-axis, it is caused by dynamics in many or all canonical coordinates.
This makes it hard to establish an analytical link for the dependence of the divergence on the system parameters (moments of inertia and electric field strength).
This is in contrast to many ``traditional'' studies of chaos in molecular system, where the chaos is related to configuration space only, allowing to define a critical energy below which chaotic regions of configuration space can not be reached~\cite{brumer74,duff76,brumer81}.

\begin{figure}
\includegraphics[width=3.375 in]{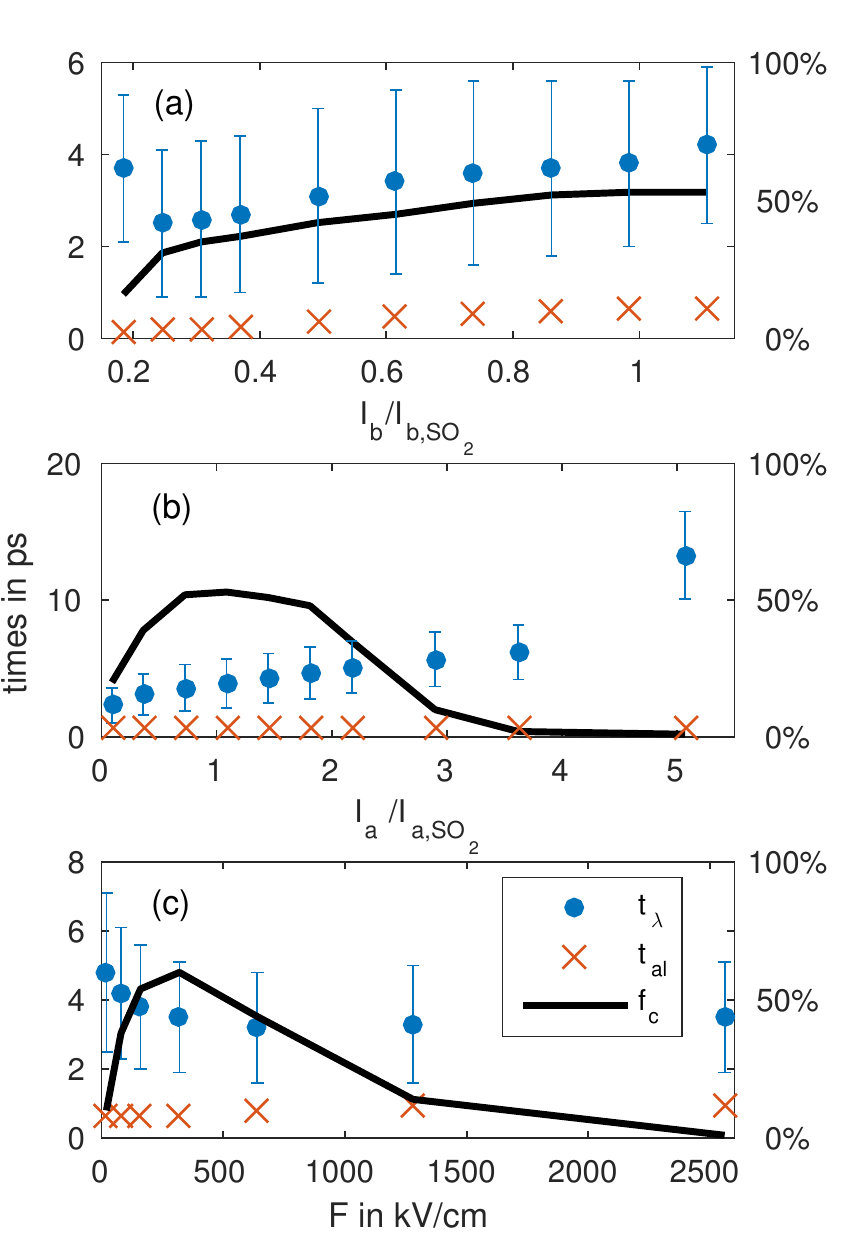}
\caption{
\label{fig.speed}
Lyapunov times $t_{\lambda}$ (circles) and time $t_{\mathrm{al}}$ (crosses) of the first alignment peak, for classical SO$_2$-like molecules in a static electric field after interaction with a short laser pulse.
Shown are also the percentage of chaotic trajectories (black line and right side ordinate).
In panel (a), the results are shown as a function of the moment of inertia $I_b$ of the $b$-axis, in panel (b) as a function of the moment of inertia $I_a$ of the $a$-axis, and in panel (c) as a function of the strength $F$ of the static field.
}
\end{figure}

Lacking an analytical connection between the system parameters and the chaotic dynamics other than the equations of motion themselves, we revert to numerical simulations.
Specifically, we calculated the time of the alignment peak and the ensemble-averaged Lyapunov time after the laser pulse as a function of different system parameters.
The base system are SO$_2$ molecules at 10~K, in a static field of 150~kV/cm, interacting with a laser pulse of 50~fs duration and 50~TW/cm$^2$ peak intensity (i.e., the parameters of the partly chaotic system from the preceding section).
The results are shown in Fig.~\ref{fig.speed}, where circles show the Lyapunov times, squares show the alignment time, and the black line displays the fraction of molecules that end up with a chaotic trajectory after the laser pulse.

The calculated dependence on a change of the moment of inertia along the $b$-axis, $I_b$, which is bounded by $I_a$ and $I_c$, is shown in Fig.~\ref{fig.speed}~(a), where it is clear that the Lyapunov time remains an order of magnitude above the alignment time for all possible values of $I_b$. 

A more promising parameter is $I_a$, for two reasons:
Its lower bound is zero, and the alignment of the $a$-axis can be expected to be independent of a rotation around the $a$-axis.
Thus, in principle, by decreasing $I_a$ sufficiently we could reduce the Lyapunov time below the alignment time.
The results are shown in Fig.~\ref{fig.speed}~(b), and indeed $t_
{\mathrm{al}}$ remains constant whereas $t_{\lambda}$ decreases with $I_a$.
However, the smaller that $I_a$ becomes, the more the molecule's shape becomes linear -- with the result that the fraction of chaotic trajectories decreases to zero, long before $t_{\lambda}\approx t_{\mathrm{al}}$.

As a third test, we varied the strength $F$ of the static field.
Again, as shown in Fig.~\ref{fig.speed}~(c), the alignment time is almost independent from $F$, whereas $t_{\lambda}$ decreases monotonically.
However, in this case as well the window for which the dynamics is chaotic is too small to achieve $t_{\mathrm{al}}\approx t_{\lambda}$.

From these investigations we conclude that laser-assisted alignment is -- at least for all investigated scenarios -- robust against rotational chaos.

\section{\label{sec.conclusion}Discussion and conclusion}

We have investigated laser-induced molecular alignment in a system with rotational chaos:
SO$_2$ molecules in a static electric field, aligned by a femtosecond non-resonant laser pulse.
The control scheme was found to be robust against rotational chaos.
Both the transient alignment peak shortly after the laser pulse as well as the time-averaged alignment were essentially unaffected by the additional static field, although investigations on the classical systems showed that a large fraction of the phase space displayed chaotic dynamics.
The only affected part of the control scheme were the quantum mechanical revivals, which were lost when the chaos became too strong.

Using numerical simulations we showed that the alignment dynamics were faster than the chaos-induced exponential divergence for all investigated scenarios.
This indicates that alignment dynamics are inherently faster, and therefore robust against the chaos.
If this is indeed the case, we would expect that excitation schemes that operate on longer time-scales are more affected by chaos.
One such long-time scheme is the optical centrifuge~\cite{karczmarek99,villeneuve00,yuan11,korobenko13}, which uses pulses that are typically tens of picoseconds long.
In a recent experiment~\cite{korobenko16}, Korobenko~\textit{et al.} achieved planar alignment of SO$_2$ molecules by an optical centrifuge.
Since the centrifuge pulse itself could induce chaotic dynamics, chaos may have played a role in that experiment.
A study examining this possibility is therefore motivated.

In many chaotic systems previously studied in molecular dynamics, the chaos is generated by the coupling of coordinates via the potential.
By determining the regions of configuration space for which the potential causes chaotic divergence, one can define a critical energy below which the chaotic parts of the potential can not be reached and the dynamics remain regular~\cite{duff76}.
For a rotational system, the dynamics is more complicated, with momenta being coupled as well.
Hence the chaotic divergence is likely generated in phase space rather than  configuration space.
For a future study, it would thus be interesting to find an equivalent of the critical energy for phase space.

\acknowledgements

This work was supported by NSERC and a fellowship within the Postdoc-Program of the German Academic Exchange Service (DAAD).




%

\end{document}